\documentclass[preprint]{revtex4}
\usepackage{graphicx,float}
\usepackage{amssymb,amsmath}

\begin{document}
\title{Neutron Anomalous Magnetic Moment in Dense Magnetized Systems}

\author{Zeinab Rezaei\footnote{E-mail: zrezaei@shirazu.ac.ir}}

\affiliation{Physics Department
and Biruni Observatory, College of Sciences, Shiraz
University, Shiraz 71454, Iran}

\begin{abstract}
In this work, we calculate the neutron anomalous magnetic moment
supposing that this value can depend on the density and magnetic field of system.
We employ the lowest order constraint variation (LOCV) method and $AV_{18}$ nuclear potential to calculate the medium dependency of the neutron anomalous magnetic moment.
It is confirmed that the neutron anomalous magnetic moment
increases by increasing the density, while it decreases as the magnetic field grows.
The energy and equation of state for the system have also been investigated.

\textbf{Keywords}:neutron; anomalous magnetic moment; magnetic field
\end{abstract}
\maketitle
\section{Introduction}
High-density neutron matter and nuclear matter with strong magnetic field can be found
in the interior of neutron stars. Therefore, investigation of the nucleonic matter
with high density and strong magnetic fields are of great interest in nuclear astrophysics.
In such conditions for the density and magnetic field, the magnitude of the nucleon anomalous magnetic moments (AMM)
can be different from the free nucleon and its value can change when the physical conditions of the medium vary.
We note that having various magnitudes of the nucleon AMMs
may lead to significant consequences for the thermodynamic properties of the neutron and nuclear matter.
Accordingly, study of the dependency of nucleon AMMs
on the physical parameters of the medium, (e.g. density, magnetic field, etc.) seems necessary.

Many works have been focused on the
dependence of the nucleon AMMs on the conditions of the medium \cite{Meibner,Cheon,Saito,Frank,Lu,Smith,Horikawa,Ryu9,Ryu10,Diener,Ferrer}.
Chiral symmetry constraints on scale changes of the nucleon in a nuclear medium have
been investigated within the framework of a chiral non-linear meson theory \cite{Meibner}. It has been
shown that the isoscalar AMM of nucleon increases with increase in the density.
In the framework of the cloudy bag model and by introducing the effective masses
of mesons and nucleons, the bound nucleon AMMs have been calculated \cite{Cheon}.
It has been confirmed that the nucleon AMMs are enhanced compared
to the free ones.
Using a self-consistent quark model for nuclear matter, the variations of the masses of the non-strange vector
mesons, the hyperons, and the nucleons in the dense nuclear matter have been investigated \cite{Saito}. In this reference,
the authors have shown that
the AMM of the proton in symmetric nuclear matter increases with density.
They have also confirmed that in the bag model, the attractive scalar potential leads to the decreasing of quark mass, and
the lower component of the wave function is enhanced, leading to the increase of the AMM
of the proton and the other hadrons.
Using the ideas of color neutrality, the influence of the nuclear medium
upon the internal structure of a composite nucleon has been studied \cite{Frank}.
It has been concluded that the medium effect is an increase in the value of the AMM.
By calculating the electric and magnetic
form factors for the proton, bound in specific shell-model
orbits, it has been found that the AMM of the
bound proton is increased by the medium modifications \cite{Lu}.
They have also pointed out that this medium correction
is solely due to the change of the internal quark structure.
Chiral quark-soliton model has been employed to calculate the electromagnetic form factors of a bound proton \cite{Smith}. The results show the enhancement of the AMM.
Applying Nambu-Jona-Lasinio model to investigate the medium modifications
of the nucleon electromagnetic form factors, it has been shown that the medium
effects tend to decrease the intrinsic AMM of the proton but when combined with the enhancement of the nuclear magneton, the
spin g-factor is enhanced \cite{Horikawa}.
AMMs of hyperons in dense nuclear matter have been calculated using relativistic quark
models in which hyperons have been treated as MIT bags and the interactions have been
considered to be mediated by the exchange of scalar and vector mesons \cite{Ryu9}.
The results confirm that the magnitudes of the AMMs increase with density
for most octet baryons.
Using a quantum hadrodynamic model, the medium effects caused by density-dependent AMMs of baryons on neutron stars
under strong magnetic fields have been studied \cite{Ryu10}. It has been found that the
AMMs of nucleons can be enhanced to be larger than those of hyperons.
Strongly magnetized symmetric nuclear matter is investigated within the context of effective
baryon-meson exchange models \cite{Diener}. It has been found that by
increasing the dipole moment strength, the system becomes more
tightly bound.
The influence of the AMM on the equation of state
of charged fermions in the presence of a magnetic field has been considered \cite{Ferrer}.
In this work, the AMM has been found from the one-loop fermion self-energy.
It has been concluded that in the strong magnetic field region the AMM depends on the
Landau level. Their results show that the AMM of charged fermions have no significant
effects on the equation of state.

In addition to predict the dependency of AMM on the medium, it is important to
find the way that the physical parameters affect
the nucleon AMMs.
Many authors have explored the effects of medium on the intrinsic properties of nucleons
and how the modifications of the AMM occur.
From an analysis of the structure functions for inelastic electron scattering, it has been found
that the charge radius and the AMM of nucleons increase in $^{12}C$, due to the effect of the
nuclear medium on the quark wave functions \cite{Mulders}. It has been concluded that from the increases in the nucleon
radius, one also expects an increase of the AMM, since for massless quarks in the nucleon,
the AMM is proportional to the size of the quark wave function. It has been also found that
the AMM and radius are the best quantities from which to deduce the size of the
quark wave functions in nuclei.
Besides, it has been indicated that the proton and neutron charge radii increase with density \cite{Meibner}.
In a chiral nonlinear quark-meson theory, it has been shown that in the presence of an external baryon medium,
the proton radius increases \cite{Arriola}.
It has been argued that the
increase in the AMM tends
to cancel the effect of the increased radius \cite{Frank}.
Besides, it has been concluded that at low values of
the square of the momentum transfer, the electric form factor is suppressed and displays
an increased charge radius, while the magnetic
radius and the AMM are increased.
Moreover, it has been shown that the
electromagnetic rms radii and the AMM of the
bound proton are increased by the medium modifications \cite{Lu}.
They have found that the intrinsic AMM is
enhanced in matter because of the change in the quark structure
of the nucleon.
Using MIT bag model, it has been found that in the presence of ultra-strong magnetic fields, a nucleon either flattens or collapses in the direction transverse to the external magnetic field in the
classical or quantum mechanical picture respectively \cite{Mandal}.
According to Ref. \cite{Ryu9}, there is a big difference between
the bag properties obtained from the quark-meson coupling (QMC) and
modified quark-meson coupling (MQMC) models.
In the QMC model, the bag radius decreases as the density increases, but in the MQMC
model, the bag radius increases with density \cite{Jin}. It has been concluded that since the AMM depends on the bag radius, the
prediction of the AMM in the MQMC model will
differ from that obtained from the QMC model.
In addition, the authors of Ref.\cite{Ryu10} believe that the medium
effects due to density-dependent AMMs
are larger in higher magnetic fields.

In our previous study, we have calculated the magnetic properties of neutron matter in the presence of strong magnetic fields using the lowest order constraint variation (LOCV) method assuming that the neutron AMM is not affected by the medium \cite{Bordbar11}. In the present work, we are interested in the medium dependency of the neutron magnetic moment as well as the
properties of  magnetized neutron matter with the medium dependent AMM using the LOCV method applying $AV_{18}$ nuclear potential.

\section{LOCV formalism for magnetized neutron matter with the medium dependent anomalous magnetic moment}
\label{sec:1}
We start with a pure homogeneous system of spin polarized neutrons with the spin-up $(+)$ and
spin-down $(-)$ states. The number densities of spin-up and spin-down neutrons are shown
by $\rho^{(+)}$ and $\rho^{(-)}$, respectively. The spin polarization parameter
$\delta=\frac{\rho^{(+)}-\rho^{(-)}}{\rho}$, is introduced
where $\rho=\rho^{(+)}+\rho^{(-)}$ is the total density of system.
We take the uniform magnetic field along the $z$ direction,
$B=B\widehat{k}$, which leads the spin up and down particles corresponding to
parallel and antiparallel spins with respect to the magnetic field.
In this work, LOCV method is applied to calculate the energy of the system as follows.

We consider a trial many-body wave function of the form
\begin{eqnarray}
     \psi=F\phi,
 \end{eqnarray}
where $\phi$ is the uncorrelated ground-state wave function of
$N$ independent neutrons, and $F$ is a proper $N$-body correlation
function. Jastrow approximation \cite{Jastrow} is employed in which $F$ can be
replaced by
\begin{eqnarray}
    F=S\prod _{i>j}f(ij),
 \end{eqnarray}
where $S$ is a symmetrizing operator. We consider a cluster expansion of the
energy functional up to the two-body term,
 \begin{eqnarray}\label{tener}
E([f])=\frac{1}{N}\frac{\langle\psi|H|\psi\rangle}
{\langle\psi|\psi\rangle}=E _{1}+E _{2}\cdot
\end{eqnarray}
The one-body term, $E_{1}$, for magnetized neutron matter is given by
\begin{eqnarray}
 \label{oneterm}
E_{1}=\sum_{i=+,-}\frac{3}{5}\frac{\hbar^{2}k_{F}^
{(i)^{2}}}{2m}\frac{\rho^{(i)}}{\rho}-\mu_{dep} B \delta,
\end{eqnarray}
where $k_{F}^{(i)}=(6\pi^{2}\rho^{(i)})^{\frac{1}{3}}$ is the
Fermi momentum of a neutron with spin projection $i$ and $\mu_{dep}$ is the value of neutron AMM that can depend on the density and magnetic field of the system.
We define the parameter $r_{\mu}=\mu_{dep}/{\mu_{n}}$ in which ${\mu_{n}= -1.9130427(5)}$ is the AMM of the free neutron.
The dimensionless parameter $r_{\mu}$ quantifies the medium dependent neutron AMM.
The value $r_{\mu}=1$ corresponds to the AMM of the free neutron.
The two-body energy, $E_{2}$, is as follows,
\begin{eqnarray}
E_{2}&=&\frac{1}{2N}\sum_{ij} \langle ij\left| \nu(12)\right|
    ij-ji\rangle,
\end{eqnarray}
where
$$\nu(12)=-\frac{\hbar^{2}}{2m}[f(12),[\nabla
_{12}^{2},f(12)]]+f(12)V(12)f(12).$$
In the above equation, $f(12)$
and $V(12)$ are the two-body correlation function and nuclear
potential, respectively. In order to calculate the energy of neutron matter, we employ the
$AV_{18}$ two-body nuclear potential \cite{Wiringa},
\begin{eqnarray}
V(12)&=&\sum^{18}_{p=1}V^{(p)}(r_{12})O^{(p)}_{12}
\end{eqnarray}
where $O^{(p)}_{12}$ shows the operators in $AV_{18}$ potential \cite{Wiringa}.
In our formalism, we consider the two-body correlation function, $f(12)$, as follows \cite{Owen},
\begin{eqnarray}
f(12)&=&\sum^3_{k=1}f^{(k)}(r_{12})P^{(k)}_{12},
\end{eqnarray}
where
\begin{eqnarray}
P^{(k=1-3)}_{12}&=&(\frac{1}{4}-\frac{1}{4}O_{12}^{(2)}),\
(\frac{1}{2}+\frac{1}{6}O_{12}
^{(2)}+\frac{1}{6}O_{12}^{(5)}),\nonumber\\&&
(\frac{1}{4}+\frac{1}{12}O_{12}^{(2)}-\frac{1}{6}O_{12}^{(5)}).
\end{eqnarray}
The operators $O_{12}^{(2)}$ and $O_{12}^{(5)}$ are given in \cite{Wiringa}.
Using the mentioned two-body correlation function and
potential, after doing some algebra,
the two-body energy is obtained as follows,
\begin{eqnarray}\label{ener2}
    E_{2} &=& \frac{2}{\pi ^{4}\rho }\left( \frac{\hbar^{2}}{2m}\right)
    \sum_{JLSS_{z}}\frac{(2J+1)}{2(2S+1)}[1-(-1)^{L+S+1}] \nonumber
\\&& \times\left| \left\langle
\frac{1}{2}\sigma _{z1}\frac{1}{2}\sigma _{z2}\mid
SS_{z}\right\rangle \right| ^{2}  \int_0^{\infty} dr\left\{\left [{f_{\alpha
}^{(1)^{^{\prime }}}}^{2}{a_{\alpha
}^{(1)}}^{2}(r,\rho^{(i)}) \right.\right. \nonumber \\&&\left.\left.
 +\frac{2m}{\hbar^{2}}(\{V_{c}-3V_{\sigma } +V_{\tau }-3V_{\sigma
\tau }+2(V_{T}-3V_{\sigma T }) \right.\right. \nonumber \\&&\left.\left.-2V_{\tau z}\}{a_{\alpha
}^{(1)}}^{2}(r,\rho^{(i)})
+[V_{l2}-3V_{l2\sigma } +V_{l2\tau }-3V_{l2\sigma \tau }] \right.\right. \nonumber \\&&\left.\left.{\times c_{\alpha
}^{(1)}}^{2}(r,\rho^{(i)}))(f_{\alpha }^{(1)})^{2}\right ]
+\sum_{k=2,3}\left[ {f_{\alpha }^{(k)^{^{\prime }}}}^{2}{a_{\alpha
}^{(k)}}^{2}(r,\rho^{(i)})\right.\right. \nonumber \\&&\left. \left.
+\frac{2m}{\hbar^{2}}( \{V_{c}+V_{\sigma }+V_{\tau } +V_{\sigma \tau
}+(-6k+14)(V_{t\tau} \right.\right. \nonumber \\&&\left.\left.+V_{t})-(k-1)(V_{ls\tau }+V_{ls})+2[V_{T}+V_{\sigma T }\right.\right. \nonumber \\&&\left.\left.+(-6k+14)V_{tT}-V_{\tau
z}]\}{a_{\alpha }^{(k)}}^{2}(r,\rho^{(i)}) +[V_{l2}+V_{l2\sigma }\right.\right. \nonumber \\&&\left.\left. +V_{l2\tau }+V_{l2\sigma \tau
}]{c_{\alpha }^{(k)}}^{2}(r,\rho^{(i)})+[V_{ls2}+V_{ls2\tau }]\right.\right. \nonumber \\&&\left.\left. {\times d_{\alpha
}^{(k)}}^{2}(r,\rho^{(i)})) {f_{\alpha }^{(k)}}^{2}\right ]+\frac{2m}{\hbar^{2}}\{V_{ls}+V_{ls\tau
}-2(V_{l2}+V_{l2\sigma }\right. \nonumber \\&&\left.+V_{l2\sigma \tau } +V_{l2\tau })-3(V_{ls2}
+V_{ls2\tau })\}b_{\alpha }^{2}(r,\rho^{(i)})f_{\alpha }^{(2)}f_{\alpha
}^{(3)}\right. \nonumber \\&&\left. +\frac{1}{r^{2}}(f_{\alpha
}^{(2)} -f_{\alpha }^{(3)})^{2}b_{\alpha }^{2}(r,\rho^{(i)})\right\},
 \end{eqnarray}
with the definition for $\alpha=\{J,L,S,S_z\}$. The coefficient  ${a_{\alpha
}^{(1)}}^{2}$, etc., are as follows,
\begin{eqnarray}\label{a1}
     {a_{\alpha }^{(1)}}^{2}(x,\rho)=x^{2}I_{L,S_{z}}(x,\rho),
 \end{eqnarray}
\begin{eqnarray}
     {a_{\alpha }^{(2)}}^{2}(x,\rho)=x^{2}[\beta I_{J-1,S_{z}}(x,\rho)
     +\gamma I_{J+1,S_{z}}(x,\rho)],
 \end{eqnarray}
\begin{eqnarray}
           {a_{\alpha }^{(3)}}^{2}(x,\rho)=x^{2}[\gamma I_{J-1,S_{z}}(x,\rho)
           +\beta I_{J+1,S_{z}}(x,\rho)],
      \end{eqnarray}
\begin{eqnarray}
     b_{\alpha }^{2}(x,\rho)=x^{2}[\beta _{23}I_{J-1,S_{z}}(x,\rho)
     -\beta _{23}I_{J+1,S_{z}}(x,\rho)],
 \end{eqnarray}
\begin{eqnarray}
         {c_{\alpha }^{(1)}}^{2}(x,\rho)=x^{2}\nu _{1}I_{L,S_{z}}(x,\rho),
      \end{eqnarray}
\begin{eqnarray}
        {c_{\alpha }^{(2)}}^{2}(x,\rho)=x^{2}[\eta _{2}I_{J-1,S_{z}}(x,\rho)
        +\nu _{2}I_{J+1,S_{z}}(x,\rho)],
 \end{eqnarray}
\begin{eqnarray}
       {c_{\alpha }^{(3)}}^{2}(x,\rho)=x^{2}[\eta _{3}I_{J-1,S_{z}}(x,\rho)
       +\nu _{3}I_{J+1,S_{z}}(x,\rho)],
 \end{eqnarray}
\begin{eqnarray}
     {d_{\alpha }^{(2)}}^{2}(x,\rho)=x^{2}[\xi _{2}I_{J-1,S_{z}}(x,\rho)
     +\lambda _{2}I_{J+1,S_{z}}(x,\rho)],
 \end{eqnarray}
\begin{eqnarray}\label{d2}
     {d_{\alpha }^{(3)}}^{2}(x,\rho)=x^{2}[\xi _{3}I_{J-1,S_{z}}(x,\rho)
     +\lambda _{3}I_{J+1,S_{z}}(x,\rho)],
 \end{eqnarray}
with
\begin{eqnarray}
          \beta =\frac{J+1}{2J+1},\ \gamma =\frac{J}{2J+1},\
          \beta _{23}=\frac{2J(J+1)}{2J+1},
 \end{eqnarray}
\begin{eqnarray}
       \nu _{1}=L(L+1),\ \nu _{2}=\frac{J^{2}(J+1)}{2J+1},
      \end{eqnarray}
\begin{eqnarray}
         \nu _{3}=\frac{J^{3}+2J^{2}+3J+2}{2J+1},
      \end{eqnarray}
\begin{eqnarray}
     \eta _{2}=\frac{J(J^{2}+2J+1)}{2J+1},\ \eta _{3}=
     \frac{J(J^{2}+J+2)}{2J+1},
 \end{eqnarray}
\begin{eqnarray}
     \xi _{2}=\frac{J^{3}+2J^{2}+2J+1}{2J+1},\
     \xi _{3}=\frac{J(J^{2}+J+4)}{2J+1},
 \end{eqnarray}
\begin{eqnarray}
     \lambda _{2}=\frac{J(J^{2}+J+1)}{2J+1},\
     \lambda _{3}=\frac{J^{3}+2J^{2}+5J+4}{2J+1}.
 \end{eqnarray}
In the above equations, the terms $a_{\alpha }^{(i)}$, $b_{\alpha }$, $c_{\alpha }^{(i)}$, and $d_{\alpha }^{(i)}$ have dimension $L^{-2}$, and $x$ has dimension $L$. In addition, $I(x,\rho)$ with dimension
$L^{-6}$ is given by
\begin{eqnarray}
       I_{J,S_{z}}(x,\rho)=\int_0^{\infty} dq\ q^2 P_{S_{z}}(q) J_{J}^{2}(xq)\cdot
 \end{eqnarray}
In the last equation, the parameter $q$ has dimension $L^{-1}$,  $J_{J}(xq)$ is the spherical Bessel function and $P_{S_{z}}(q)$ is defined as
\begin{eqnarray}
P_{S_{z}}(q)&=&\frac{2}{3}\pi[(k_{F} ^{\sigma_{z1}})^{3}+(k_{F}
^{\sigma_{z2}})^{3}-\frac{3}{2}((k_{F} ^{\sigma_{z1}})^{2}+(k_{F}
^{\sigma_{z2}})^{2})q\nonumber\\ &-&\frac{3}{16}((k_{F}
^{\sigma_{z1}})^{2}-(k_{F} ^{\sigma_{z2}})^{2})^{2}q^{-1}+q^{3}]
 \end{eqnarray}
for                 $\frac{1}{2}|k_{F} ^{\sigma_{z1}}-k_{F}
^{\sigma_{z2}}|<q<\frac{1}{2}|k_{F} ^{\sigma_{z1}}+k_{F}
 ^{\sigma_{z2}}|$,
\begin{eqnarray}
P_{S_{z}}(q)=\frac{4}{3}\pi min((k_{F} ^{\sigma_{z1}})^3,(k_{F}
^{\sigma_{z2}})^3)
 \end{eqnarray}
for  $q<\frac{1}{2}|k_{F} ^{\sigma_{z1}}-k_{F}
 ^{\sigma_{z2}}|$, and
 \begin{eqnarray}
       P_{S_{z}}(q)=0
 \end{eqnarray}
for  $q>\frac{1}{2}|k_{F} ^{\sigma_{z1}}+k_{F}
 ^{\sigma_{z2}}|$, where $\sigma_{z1}$ or
 $\sigma_{z2}= +1,-1$ for spin up and down,
 respectively.
In the next step, the two-body energy is minimized with
respect to the
variations in the function $f_{\alpha}^{(i)}$ subject to the
normalization constraint
\cite{Bordbar57},
\begin{eqnarray}\label{213}
        \frac{1}{N}\sum_{ij}\langle ij\left| h_{S_{z}}^{2}
        -f^{2}(12)\right|
ij\rangle _{a}=0,
 \end{eqnarray}
where in the case of magnetized neutron matter, the function
$h_{S_{z}}(r)$ is defined as follows,
\begin{eqnarray}
h_{S_{z}}(r)&=& \left\{\begin{array}{ll} \left[ 1-9\left(
\frac{J_{J}^{2}
(k_{F}^{(S_{z})}r)}
{k_{F}^{(S_{z})}r}\right) ^{2}\right] ^{-1/2} &;~~ S_{z}=\pm1   \\ \\
1 &;~~ S_{z}= 0.
\end{array}
\right.
\end{eqnarray}
The minimization of the two-body cluster energy leads a set of
Euler-Lagrange differential equations with the forms,
\begin{eqnarray}
&g_\alpha^{(1)^{\prime\prime}}-\{\frac{a_\alpha^{(1)^{\prime\prime}}}
{a_\alpha^{(1)}}+\frac{m}{\hbar^2}
[V_c-3V_\sigma+V_\tau-3V_{\sigma\tau}\nonumber\\&
+2(V_T-3V_{\sigma T})-2V_{\tau z}+\lambda
]+\frac{m}{\hbar^2}(V_{l2}\nonumber\\&-3V_{{l2}\sigma}+
V_{{l2}\tau}-3V_{{l2}\sigma\tau})
\frac{c_\alpha^{(1)^2}}{a_\alpha^{(1)^2}}\}g_\alpha^{(1)}=0,
\end{eqnarray}
\begin{eqnarray}
&g_\alpha^{(2)^{\prime\prime}}-\{\frac{a_\alpha^{(2)^{\prime\prime}}}
{a_\alpha^{(2)}}+\frac{m}{\hbar^2} [V_c+V_\sigma+2V_t-V_{{ls}}
+V_\tau+V_{\sigma\tau}\nonumber\\&+2V_{t\tau}
-V_{{ls}\tau}+2(V_T+V_{\sigma
T}+2V_{tT})-2V_{\tau z}+\lambda]
+\frac{m}{\hbar^2}[V_{l2}\nonumber\\&+V_{{l2}\sigma}
+V_{{l2}\tau}+V_{{l2}\sigma\tau}]\times
\frac{c_\alpha^{(2)^2}}{a_\alpha^{(2)^2}}
+\frac{m}{\hbar^2}[V_{{ls}2}+
V_{{ls}2\tau}]\frac{d_\alpha^{(2)^2}}{a_\alpha^{(2)^2}}\nonumber\\&
+\frac{b_\alpha^2}{r^2a_\alpha^{(2)^2}}\}g_\alpha^{(2)}
+\{\frac{1}{r^2}-\frac{m}{2\hbar^2}[V_{ls}-
2V_{l2}-2V_{{l2}\sigma}-3V_{{ls}2}\nonumber\\&
+V_{{ls}\tau}-2V_{{l2}\tau}-2V_{{l2}\sigma\tau}-
3V_{{ls}2\tau}]\}
\frac{b_\alpha^2}{a_\alpha^{(2)}a_\alpha^{(3)}}g_\alpha^{(3)}=0,
\end{eqnarray}
\begin{eqnarray}
&g_\alpha^{(3)^{\prime\prime}}-\{\frac{a_\alpha^{(3)^{\prime\prime}}}
{a_\alpha^{(3)}}+\frac{m}{\hbar^2} [V_c+V_\sigma-4V_t-2V_{ls}
+V_\tau+V_{\sigma\tau}\nonumber\\&-4V_{t\tau}-2V_{{ls}\tau}
+2(V_T+V_{\sigma T}-4V_{tT})-2V_{\tau z}+\lambda]\nonumber\\&
+\frac{m}{\hbar^2}[V_{l2}+V_{{l2}\sigma}+V_{{l2}\tau}
+V_{{l2}\sigma\tau}]
\frac{c_\alpha^{(3)^2}}{a_\alpha^{(3)^2}}
+\frac{m}{\hbar^2}[V_{{ls}2}+V_{{ls}2\tau}]\frac
{d_\alpha^{(3)^2}}{a_\alpha^{(3)^2}}\nonumber\\&
+\frac{b_\alpha^2}{r^2a_\alpha^{(2)^2}}\}g_\alpha^{(3)}
+\{\frac{1}{r^2}-\frac{m}{2\hbar^2}[V_{ls}-2V_{l2}-
2V_{{l2}\sigma}-3V_{{ls}2}\nonumber\\&
+V_{{ls}\tau}-2V_{{l2}\tau}-2V_{{l2}\sigma\tau}
-3V_{{ls}2\tau}]\}
\frac{b_\alpha^2}{a_\alpha^{(2)}a_\alpha^{(3)}}g_\alpha^{(2)}=0,
\end{eqnarray}
where
\begin{eqnarray}
g_\alpha^{(i)}(r)=f_\alpha^{(i)}(r)a_\alpha^{(i)}(r).
\end{eqnarray}
In the above equations, the primes denote differentiation with
respect to r and  the Lagrange multiplier
$\lambda$ is associated with the normalization constraint, Eq.
(\ref{213}).
Solving these differential equations leads to the results
for the  correlation functions, the two-body energy, and the total energy
per particle of the system.

\section{Results and discussion }\label{NLmatchingFFtex}
\begin{figure*}
\includegraphics{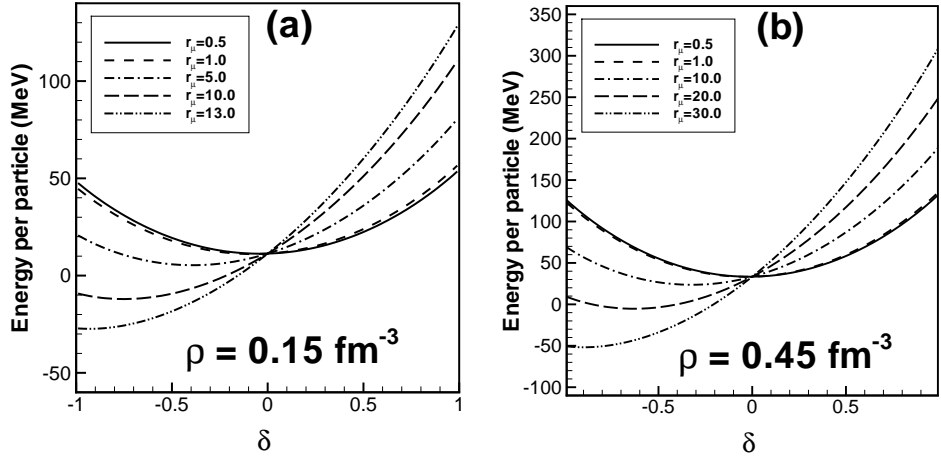}
\caption{Energy per particle versus the spin polarization parameter
at different values of dimensionless AMM, $r_{\mu}$, at $B=10^{18}\ G$.}
\label{de1}
\end{figure*}
\begin{figure*}
\includegraphics{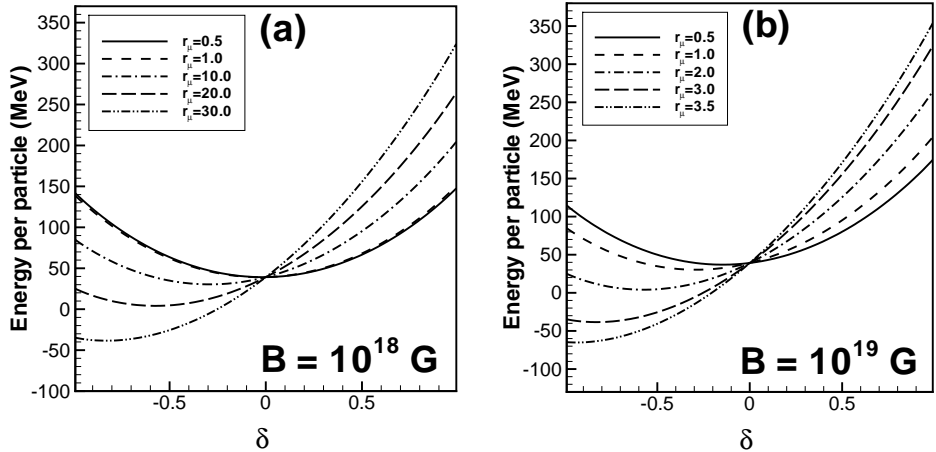}
\caption{Energy per particle versus the spin polarization parameter
at different values of dimensionless AMM, $r_{\mu}$, at $\rho=0.5\ fm^{-3}$.}
\label{de2}
\end{figure*}

Figs. \ref{de1} and \ref{de2} show the energy per particle versus the spin polarization parameter
at different values of dimensionless AMM, $r_{\mu}$. It can be seen that at
each AMM, the energy reaches a minimum at a value of the
spin polarization parameter.
The values of dimensionless AMM are acceptable that lead to an equilibrium point
with spin polarization parameter higher than $-1$, i.e. $\delta>-1$.
We can found From Figs. \ref{de1} and \ref{de2} that the energy
at the equilibrium state decreases with the increase in the dimensionless AMM.
This indicates that at high densities and magnetic fields, the neutron AMM at
which the system is stable differers from the known neutron AMM, $\mu_n$, in agreement with the result of Ref. \cite{Cheon,Frank,Lu,Smith}.
In addition, it is clear that the neutron matter with the medium dependent AMM is more
spin polarized compared to the case with $r_{\mu}=1$. It is possible to find the equilibrium state of the system
by varying the AMM. Comparing Fig. \ref{de1} a and b shows that at higher densities, the value of the
dimensionless AMM corresponding to the equilibrium state is larger than lower densities. In addition, we can see from
Fig. \ref{de2} a and b that at higher magnetic fields, the equilibrium value of the dimensionless AMM is smaller than the lower magnetic fields.
The effects of density and magnetic field on the equilibrium value of the AMM will be considered in the
following.

\begin{figure*}
\includegraphics{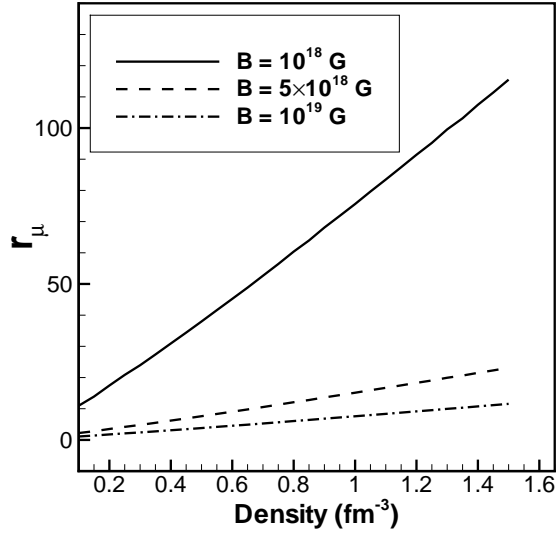}
\caption{The equilibrium value of the dimensionless
AMM versus the density at different magnetic fields.}
\label{mr}
\end{figure*}

\begin{figure*}
\includegraphics{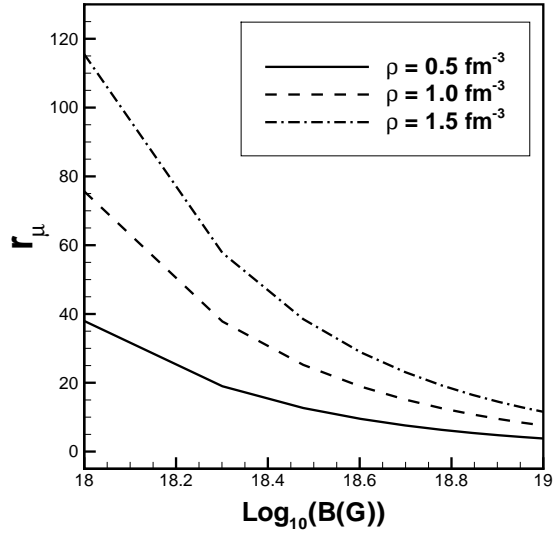}
\caption{The equilibrium value of the dimensionless
AMM versus the magnetic field at different densities.}
\label{br}
\end{figure*}

We have shown the density and magnetic field dependence of the equilibrium value of the
AMM in Figs. \ref{mr} and \ref{br}, respectively. It is clear from Fig. \ref{mr} that at each
magnetic field, the value of the dimensionless AMM increases as the density grows. This
result is in agreement with the results reported in Refs. \cite{Meibner,Saito,Ryu9}.
The enhancement of the neutron AMM can be due to the increase in the neutron
radius at higher densities \cite{Mulders,Meibner} and the change in the quark structure
of neutron \cite{Lu}.
We understand from Fig. \ref{mr} that in our model, the coupling of neutrons to the magnetic field is more significant at higher densities.
It is obvious from  Fig. \ref{mr} that the increase of the dimensionless AMM due to the density is more significant at lower magnetic fields.

Fig. \ref{br} confirms that at each density, the
dimensionless AMM decreases when the magnetic field grows.
The decrease of the AMM with the increase in the magnetic field has been also
reported in a previous work \cite{Ferrer}.
This result is expected considering the quark wave functions of the neutrons.
From the quantum mechanical point of view, strong magnetic fields result in
collapse of neutrons, and therefore the decrease in the neutron radius \cite{Mandal}.
Moreover, the AMM is proportional to the size of the quark wave function \cite{Mulders}.
Consequently, strong magnetic fields lead to the decrease in the AMM.
We see from Fig. \ref{br}  that the coupling of neutrons to the magnetic field is weaker at higher magnetic fields.
Furthermore, the effects of the density on the AMM is less significant at higher magnetic fields.

\begin{figure*}
\includegraphics{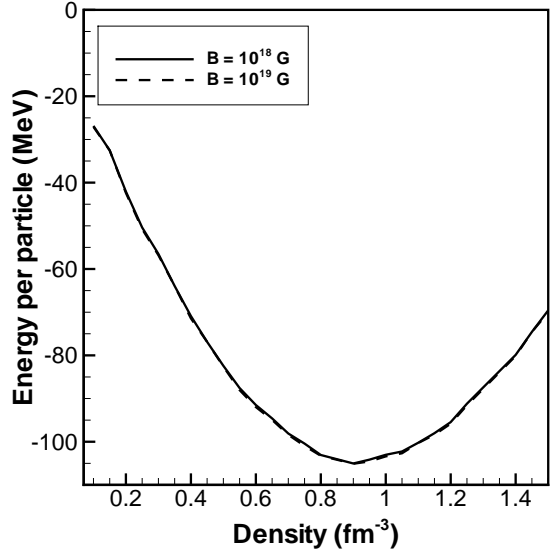}
\caption{Energy of neutron matter with the medium dependent AMM as a function of the density at different magnetic fields.} \label{fig2}
\end{figure*}
\begin{figure*}
\includegraphics{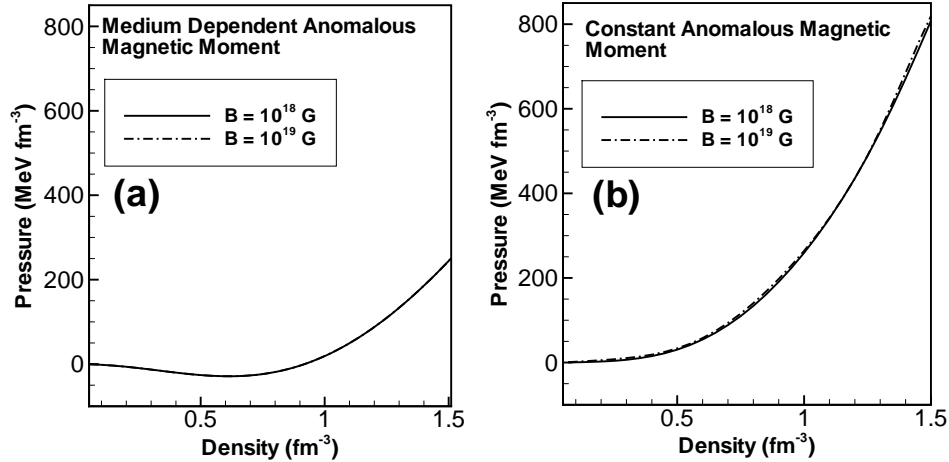}
\caption{The equation of state of neutron matter with the
 medium dependent AMM (a) and constant AMM (b) \cite{Bordbar11} at different magnetic fields.}
\label{fig4}
\end{figure*}

Fig. \ref{fig2} shows the energy of magnetized neutron matter at the equilibrium value
of the AMM versus the density for different values of the magnetic field. We can see that for each value of the magnetic field, the neutron matter is bound and has a minimum at a specific value of the density.
This bounding of the neutron matter is the result
of the strong magnetic field which affects the value of the neutron AMM.
We found that the neutron matter with the medium dependent
AMM is more bound when the magnetic field increases.
We have given the equation of state of magnetized neutron matter in Fig. \ref{fig4}.
Our results confirm that for the system with the medium dependent AMM, the equation of state is softer compared to the constant one.
It is clear from Fig. \ref{fig4} that the equation of state
is not significantly affected by the AMM in agreement with the results of a recent
 work \cite{Ferrer}.
The soft equation of state in the present case can have astrophysical consequences
related to the neutron stars. However, the influence of the other factors such as the amount of charged particles, macroscopic magnetic field distributions, and the parameterizations of the many-body forces
in magnetized neutron stars \cite{Gomes} should also be considered.

\section{Summary and Conclusions}
Applying the lowest order constraint variational method and $AV_{18}$ nuclear potential, we investigated the properties of magnetized dense neutron matter with the medium dependent AMM.
It was clarified that the neutron magnetic
moment increases with the increase in the density.
In addition, we showed that the neutron magnetic
moment decreases as the magnetic field grows.
For our system, the energy of neutron matter has a minimum value at a specific density.
The bounding of neutron matter is due to the density and magnetic field dependence of the neutron AMM.
We found that the neutron matter is more bound when the magnetic field increases.
Moreover, the equation of state of magnetized neutron matter with the medium dependent AMM
was found to be softer compared to the case with constant AMM.

\section*{Acknowledgements}
The author wishes to thank the Shiraz University Research Council.

\end{document}